\documentclass[12pt,a4paper,final]{iopart}

\usepackage{iopams}  
\usepackage{graphicx}
\usepackage[breaklinks=true,colorlinks=true,linkcolor=blue,urlcolor=blue,citecolor=blue]{hyperref}
\usepackage{braket}
\usepackage{xcolor}
\usepackage{sidecap}
\usepackage{mathrsfs}

\begin{document}

\title[]{Abnormal superfluid fraction and structural properties of electrons in $2D$ and $3D$ quantum dots: an \textit{ab initio} path-integral Monte Carlo study}

\author{Tobias Dornheim}
\address{Center for Advanced Systems Understanding (CASUS), 03581 G\"orlitz, Germany}
\address{Helmholtz-Zentrum Dresden-Rossendorf (HZDR), 01328 Dresden, Germany}
\ead{t.dornheim@hzdr.de}

\author{Yangqian Yan}
\address{Department of Physics, The Chinese University of Hong Kong, Hong Kong, China}

\begin{abstract}
We present extensive new direct path-integral Monte Carlo results for electrons in quantum dots in two and three dimensions.
This allows us to investigate the nonclassical rotational inertia (NCRI) of the system, and we find an abnormal negative superfluid fraction [Phys.~Rev.~Lett.~\textbf{112}, 235301 (2014)] under some conditions. In addition, we study the structural properties by computing a sophisticated center-two particle correlation function. Remarkably, we find no connection between the spatial structure and the NCRI, since the former can be nearly identical for Fermi- and Bose-statistics for parameters where the superfluid fraction is diverging towards negative infinity.
\end{abstract}

\vspace{2pc}

\section{Introduction\label{sec:introduction}}

The study of fermionic many-body systems at finite temperature constitutes a highly active field of research, offering a plethora of interesting physical effects. Prominent examples include ultracold atoms~\cite{Zwerger_RevModPhys_2008,Zwierlein2006} such as $^3$He~\cite{Ceperley_PRL_1992,Skoeld_1980,Panholzer_Nature_2012,Nava_PRB_2013,dornheim2021path}, which undergoes a pairing-induced superfluid phase transition at low temperature~\cite{vollhardt2013superfluid} and exhibits an intriguing, non-trivial phonon-roton-maxon dispersion relation in the spectrum of collective excitations~\cite{Panholzer_Nature_2012,Nava_PRB_2013,dornheim2021path}.
A second relevant example is given by exotic \emph{warm dense matter}~\cite{wdm_book,fortov_review,new_POP,review,dornheim_dynamic}, which naturally occurs in astrophysical objects like brown dwarfs~\cite{saumon1,becker} and giant planet interiors~\cite{Militzer_2008,manuel}, and has been predicted to manifest on the pathway towards inertial confinement fusion~\cite{hu_ICF}. 

A particularly interesting system is given by electrons in quantum dots~\cite{Reimann_RevModPhys_2002}, which are studied in the present work. First and foremost, we mention that these exhibit a number of interesting physics, like the quantum breathing mode response to a monopole excitation~\cite{BM1,BM2}, and the formation of Wigner molecules~\cite{Egger_PRL_1999}
 and even crystallization~\cite{Filinov_PRL_2001} at strong coupling. A further important research topic regarding electrons in quantum dots is the investigation of addition energies~\cite{Addition1,Addition2,Addition3}, which can be directly compared to experimental measurements~\cite{Addition4}. Moreover, we mention the formation of vortices and their dependence on the respective spatial structure~\cite{Tavernier_PRB_2004,Tavernier_PRB_2006}, and the rich interplay of the rotational inertia with a strong external magnetic field~\cite{PhysRevB.73.075301}.

From a theoretical perspective, the accurate description of all aforementioned systems requires to simultaneously take into account i) coupling effects due to the interactions, ii) thermal excitations as a result of the finite temperature, and iii) quantum degeneracy effects like Pauli blocking since identical fermions are indistinguishable. In addition, confined few-body systems~\cite{Blume_2012} additionally require to take into account the interplay between i)-iii) and iv) the external potential, which, in the present study, is modelled as a harmonic oscillator. 
Due to these challenges, electrons in quantum dots constitute a rigorous benchmark that is often used to test novel many-body methods, e.g., Refs.~\cite{Dornheim_NJP_2015,Egger_2001,Schoof_CPP_2011,Hirshberg_JCP_2020,doi:10.1063/5.0030760,Dornheim_2021,Dornheim_CPP_2019,PhysRevE.80.066702,Lyubartsev_2005,xie2021abinitio}.

In the present work, we use the direct path-integral Monte Carlo (PIMC) method~\cite{cep,dornheim_permutation_cycles} without any nodal constraints~\cite{Ceperley1991} to simulate electrons in both $2D$ and $3D$ harmonic confinements. Therefore, our simulations are afflicted with the notorious fermion sign problem (FSP)~\cite{troyer,Loh_sign_problem_PRB_1990,dornheim_sign_problem}, but are exact within the given Monte Carlo error bars, which allows for a rigorous treatment of the intriguing interplay between the effects i)-iv) mentioned above.
From a practical perspective, the PIMC method gives us direct access to the superfluid fraction of the system~\cite{Kwon_LSF_PRL_1999,dornheim_superfluid}, which exhibits a remarkable abnormal divergence towards negative infinity in the case of fermions for some conditions~\cite{Blume_PRL_2014,Dornheim_PRA_2020}. This effect was first reported by Yan and Blume~\cite{Blume_PRL_2014} for a short-range pair potential and subsequently found by Dornheim~\cite{Dornheim_PRA_2020} for quantum-dipole systems, and has been explained in terms of the symmetry of the thermal density matrix. Here, we extend these considerations to electrons in quantum dots and find analogous behaviour.
In addition, we study the structural properties of the system using a recently suggested center-two particle (C2P) correlation function~\cite{Thomsen_PRE_2015}, which efficiently takes into account the rotational symmetry of the system.
This allows us to resolve the interesting interplay between quantum statistics and the Coulomb interaction between pairs of electrons, and to explore the connection between the aforementioned abnormal superfluid fraction and the static structure.

The paper is organized as follows: In Sec.~\ref{sec:theory}, we introduce the relevant theoretical background including the model system~(\ref{sec:introduction}), the utilized PIMC simulation approach~(\ref{sec:PIMC}), the estimation of the superfluid fraction~(\ref{sec:SF_theory}), and finally the center-two particle correlation function (\ref{sec:C2P}).
Sec.~\ref{sec:results} is devoted to the presentation of our extensive new simulation results, which we start by shortly revisiting the noninteracting system in Sec.~\ref{sec:ideal}. Subsequently, we study both the superfluid fraction and the structural properties of electrons in quantum dots, focusing on the dependence on the temperature (\ref{sec:temperature}), on the system-size (\ref{sec:N}), and on the coupling strength (\ref{sec:coupling}). The paper is concluded with a concise summary and outlook in Sec.~\ref{sec:summary}.

\section{Theory\label{sec:theory}}

\subsection{Model system\label{sec:Hamiltonian}}

We consider $N$ spin-polarized electrons in both two- and three-dimensional quantum dots, which we model by a simple harmonic potential~\cite{Reimann_RevModPhys_2002}. This gives the Hamiltonian
\begin{eqnarray}\label{eq:Hamiltonian_trap}
\hat H = - \frac{1}{2} \sum_{k=1}^N \nabla_k^2 + \frac{1}{2} \sum_{k=1}^N \mathbf{\hat r}_k^2 + \sum_{k>l}^N \frac{ \lambda }{ |\mathbf{\hat r}_l - \mathbf{\hat r}_k| } \quad ,
\end{eqnarray}
where we assume oscillator units, corresponding to the characteristic length $l_0=\sqrt{\hbar/m\Omega}$ (with $\Omega$ being the trap frequency) and energy scale $E_0=\hbar\Omega$. The first term corresponds to the kinetic contribution $\hat K$ and the last two terms to the external potential and the Coulomb interaction, $\hat V_\textnormal{ext}$ and $\hat W$, respectively. Moreover, the coupling constant $\lambda$ can, in principle, be tuned in experiments~\cite{Filinov_PRB_2008}.

\subsection{Path-integral Monte Carlo\label{sec:PIMC}}

We use the direct path-integral Monte Carlo method~\cite{Berne_JCP_1982,Takahashi_Imada_PIMC_1984}, see Ref.~\cite{cep} for an extensive review article. In a nutshell, the basic idea behind PIMC is to stochastically sample the thermal density matrix in the canonical ensemble
\begin{eqnarray}\label{eq:density_matrix}
\rho(\mathbf{R}_1,\mathbf{R}_2,\beta) = \bra{\mathbf{R}_1} e^{-\beta\hat H} \ket{\mathbf{R}_2}\ ,
\end{eqnarray}
where $\hat H$ is the Hamiltonian, $\beta=1/k_\textnormal{B}T$ the inverse temperature, and $\mathbf{R}=(\mathbf{r}_1,\dots,\mathbf{r}_N)^T$ contains the coordinates of all $N$ particles. While a direct evaluation of Eq.~(\ref{eq:density_matrix}) is not possible as different contributions to the full Hamiltonian do not commute, one makes use of a Trotter decomposition~\cite{trotter} into $P$ terms that are each given at $P$-times the original temperature. The resulting scheme becomes exact in the limit of large $P$ and the respective convergence has been carefully checked; see Ref.~\cite{brualla_JCP_2004} for a detailed discussion of different factorization schemes. The PIMC method thus allows for the quasi-exact evaluation of thermal expectation values 
\begin{eqnarray}\label{eq:expectation}
\braket{\hat A} = \textnormal{Tr}\left(
 e^{-\beta\hat H} \hat A
\right)\
\end{eqnarray}
within the given Monte Carlo error bar, which scales as $\Delta A\sim 1/\sqrt{N_\textnormal{MC}}$ with the number of Monte Carlo samples $N_\textnormal{MC}$.
In practice, we use a canonical adaption~\cite{mezza} of the worm algorithm presented by Boninsegni \textit{et al.}~\cite{boninsegni1,boninsegni2}.

An additional obstacle regarding the PIMC simulations of fermions (i.e., particles obeying Fermi-Dirac statistics such as electrons or protons) is given by the anti-symmetry of Eq.~(\ref{eq:density_matrix}) under the exchange of particle coordinates~\cite{dornheim_permutation_cycles}. First and foremost, negative terms cannot be interpreted as a probability distribution, and the expectation value from Eq.~(\ref{eq:expectation}) gets modified to~\cite{dornheim_sign_problem}
\begin{eqnarray}\label{eq:ratio}
\braket{\hat A} = \frac{\braket{\hat A\hat S}'}{\braket{\hat S}'}\ ,
\end{eqnarray}
where $\braket{\dots}'$ denotes the expectation value computed from the absolute values of the density matrix, and $\hat S$ measures the respective sign. In particular, the denominator of Eq.~(\ref{eq:density_matrix}) is called the \emph{average sign} and constitutes a straightforward measure for the amount of cancellation within the fermionic PIMC simulation.
This can easily be seen by considering the first-order term to the relative statistical uncertainty~\cite{binder},
\begin{eqnarray}
\frac{\Delta A}{A}\sim \frac{1}{S\sqrt{N_\textnormal{MC}}}\ ,
\end{eqnarray}
which directly implies that the Monte Carlo error bar increases when the sign gets small. In fact, $S$ exponentially decreases both with the inverse temperature $\beta$ and the system size $N$, which can be directly translated to $\Delta A$ as 
\begin{eqnarray}\label{eq:FSP}
\frac{\Delta A}{A} \sim \frac{e^{\beta N \Delta f}}{\sqrt{N_\textnormal{MC}}}\ ,
\end{eqnarray}
where $\Delta f$ denotes the difference in the free energy density of the original ($\braket{\dots}$) and the modified ($\braket{\dots}'$) systems. 
Eq.~(\ref{eq:FSP}) is the root of the infamous fermion sign problem~\cite{dornheim_sign_problem}, as the exponential increase of $\braket{A}$ can only be compensated by increasing the compute time as $1/\sqrt{N_\textnormal{MC}}$, which quickly becomes infeasible. 
In fact, Eq.~(\ref{eq:FSP}) constitutes the central computational bottleneck in our simulations and limits the feasible system size to $N\leq10$.

\subsection{Superfluid fraction\label{sec:SF_theory}}

The superfluid fraction of a finite system is typically defined in terms of its nonclassical rotational inertial (NCRI), i.e., its response to an infinitesimal rotation. More specifically, one makes use of the Landau two-fluid model, where the total density is decomposed into a \emph{normal} part reacting to the rotation, and a \emph{superfluid} part that remains unaffected,
\begin{eqnarray}
n = n_\textnormal{n} + n_\textnormal{sf}\ .
\end{eqnarray}
The corresponding \emph{superfluid fraction} is then defined as the ratio of the superfluid to the total density,
\begin{eqnarray}
\gamma_\textnormal{sf} = \frac{n_\textnormal{sf}}{n} = 1 - \frac{I}{I_\textnormal{cl}}\ ,
\end{eqnarray}
with $I$ and $I_\textnormal{cl}$ denoting the actual and classical moments of inertia.
Within the path-integral picture, the former can be straightforwardly estimated as~\cite{Sindzingre_PRL_1991}
\begin{eqnarray}\label{eq:area_estimator}
\gamma_\textnormal{sf} = \frac{4m^2 \braket{A_z^2} }{\beta\hbar^2 I_\textnormal{cl}} \quad ,
\end{eqnarray}
which is often being referred to as the area estimator in the respective literature.
In particular, Eq.~(\ref{eq:area_estimator}) depends on the area that is enclosed by the paths in the plane perpendicular to the rotational axis,
\begin{eqnarray}\label{eq:area}
\mathbf{A} = \frac{1}{2} \sum_{k=1}^N\sum_{i=1}^P \left( 
\mathbf{r}_{k,i} \times \mathbf{r}_{k,i+1}
\right) \quad .
\end{eqnarray}
For completeness, we note that for an ideal (i.e., noninteracting) system, $\gamma_\textnormal{sf}$ can be computed semi-analytically~\cite{krauth2006statistical,Blume_PRL_2014}, see Ref.~\cite{Dornheim_PRA_2020} for the corresponding formulas.

In addition, we note that a non-zero superfluid fraction as it is defined in Eq.~(\ref{eq:area_estimator}) does not necessarily indicate the usual phenomenon of superfluidity that manifests for example as a fricitionless flow in ultracold $^4$He~\cite{cep}. In fact, superfluidity naturally emerges as a long-range order in the off-diagonal density matrix~\cite{Shi_PRB_2005}, which is not possible for the small system sizes considered in this work. Therefore, one should interpret it strictly in terms of NCRI defined above.
Still, as the term \emph{superfluid fraction} has been used throughout the literature~\cite{Blume_PRL_2014,Dornheim_PRA_2020,Filinov_PRB_2008,dornheim_superfluid}, we will stick to it here as well.

\subsection{Structural properties\label{sec:C2P}}

\begin{figure}
\hspace*{0.25\textwidth}\includegraphics[width=0.5\textwidth]{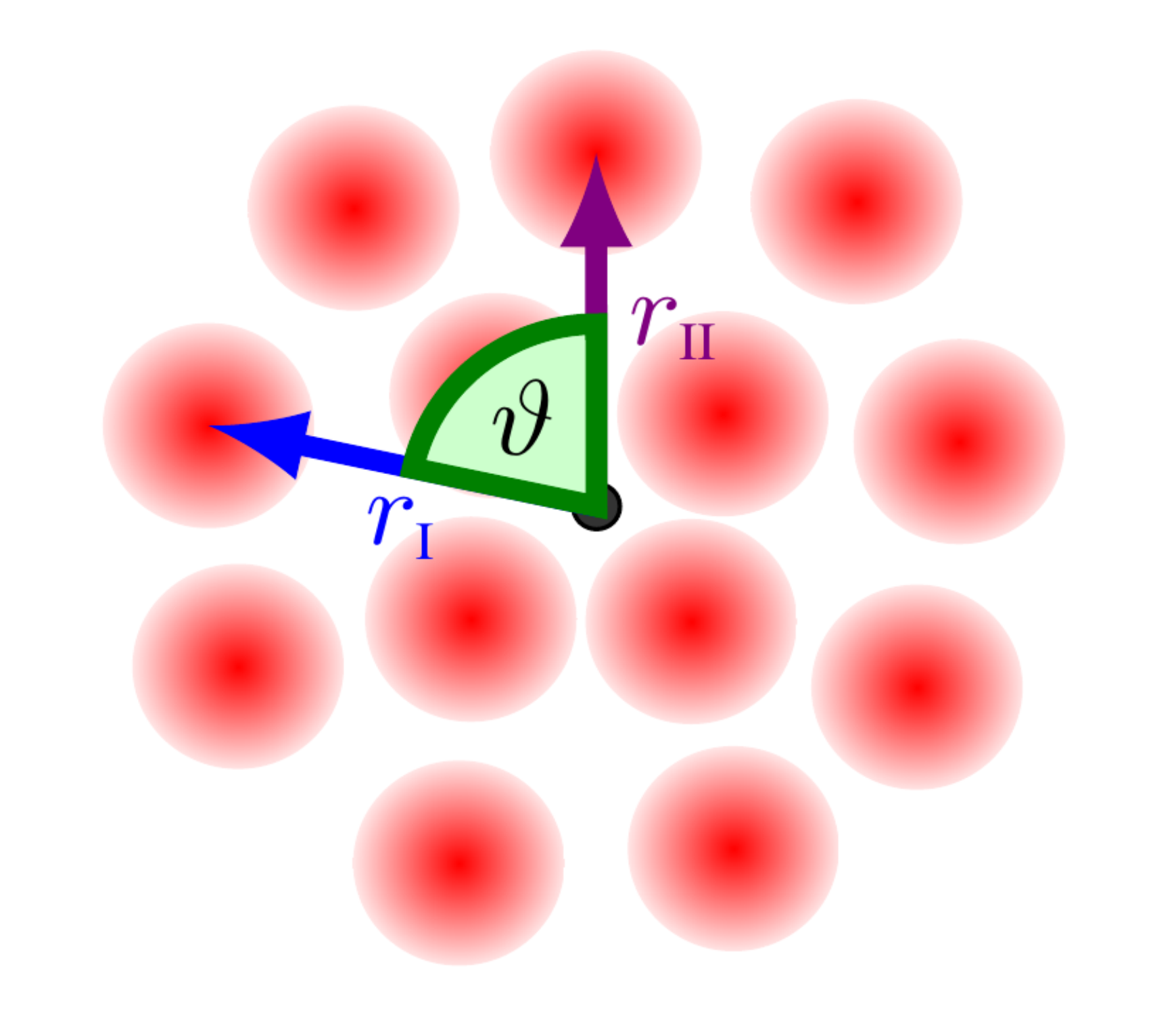}
\caption{\label{fig:c2p_illustration}
Illustration of the center-two particle correlation function $\rho_2(r_1,r_2,\vartheta)$. Due to the rotational symmetry of the Hamiltonian Eq.~(\ref{eq:Hamiltonian_trap}), the two-particle correlations only depend on the relative angle $\vartheta$ (denoted as $\alpha$ throughout this work) and the respective distances to the center of the trap.
Taken from Ref.~\cite{Dornheim_CPP_2016}. Reprinted with permission of WILEY-VCH Verlag GmbH $\&$ Co.~KGaA, Weinheim.}
\end{figure}

The second central aim of the present work is the investigation of the impact of quantum statistics on the structural properties of electrons in quantum dots.
To this end, we have implemented the center-two particle (C2P) correlation function~\cite{Thomsen_PRE_2015}, which has already been successfully applied to harmonically confined quantum systems in earlier works~\cite{Dornheim_CPP_2016,Ott2018,Dornheim_PRA_2020}.
The basic idea is illustrated in Fig.~\ref{fig:c2p_illustration}, which shows an exemplary configuration of $N=13$ particles in a $2D$ harmonic trap. Here the shaded red circles illustrate the quantum smearing of the electrons, distinguishing them from classical point particles. To characterize two-particle correlations, one can make use of the rotational symmetry of the problem and use the modified coordinates (distance to the center of the trap of particles I and II, and the angle between them, called $\vartheta$ in Fig.~\ref{fig:c2p_illustration} and $\alpha$ throughout the present work) shown in Fig.~\ref{fig:c2p_illustration} without any loss of information. More specifically, we estimate the corresponding two-particle density matrix $\rho_2(r_I,r_{II},\alpha)$ as a histogram in our PIMC simulation.

In addition, Thomsen and Bonitz~\cite{Thomsen_PRE_2015} have suggested that it would be advantageous to filter out effects that are caused by the inhomogeneous density profile $n(r)$ and not by correlation effects themselves. To this end, the actual C2P is normalized by the two-particle density matrix of a hypothetical uncorrelated system with the same radial density profile as the interacting one,
\begin{eqnarray}\label{eq:uncorrelated}
\rho^0_2(r_I,r_{II},\alpha) = \frac{N-1}{N}4\pi r_I r_{II} n(r_I) n(r_{II})\ ,
\end{eqnarray}
and is thus given by
\begin{eqnarray}\label{eq:raw_c2p}
g_\textnormal{c2p}(r_I,r_{II},\alpha) = \frac{\rho_2(r_I,r_{II},\alpha)}{\rho_2^0(r_I,r_{II},\alpha)}\ .
\end{eqnarray}
We note that Eq.~(\ref{eq:uncorrelated}) does not depend on the angle $\alpha$ by design, as noninteracting particles are not affected by their relative orientation towards each other.

While the C2P defined in Eq.~(\ref{eq:raw_c2p}) already constitutes a sophisticated tool for the investigation of spatial correlations, the visualization and analysis of a 3D quantity is often difficult. As a final step, we therefore define the integrated C2P
\begin{eqnarray}\label{eq:integrated_c2p}
g^\textnormal{int}_\textnormal{c2p}(r_I,\alpha;r_{II,min},r_{II,max}) =  \frac{\int_{r_{II,min}}^{r_{II,max}}\textnormal{d}r_{II} \rho_2(r_I,r_{II},\alpha)  }{\int_{r_{II,min}}^{r_{II,max}}\textnormal{d}r_{II} \rho_2^0(r_I,r_{II},\alpha) }\ ,
\end{eqnarray}
which can be interpreted in the following way: Eq.~(\ref{eq:integrated_c2p}) constitutes a measure for the probability to find one particle at distance $r_I$ to the center of the trap, with an angular difference of $\alpha$ to a second particle somewhere in the interval $r_{II,min}\leq r_{II} \leq r_{II,max}$, where the min and max values are typically chosen as the limits of a shell in the density profile $n(r)$.

All results for the C2P shown in this work have been obtained according to Eq.~(\ref{eq:integrated_c2p}).

\section{Results\label{sec:results}}

\subsection{Noninteracting system\label{sec:ideal}}

\begin{figure}\centering
\includegraphics[width=0.99\textwidth]{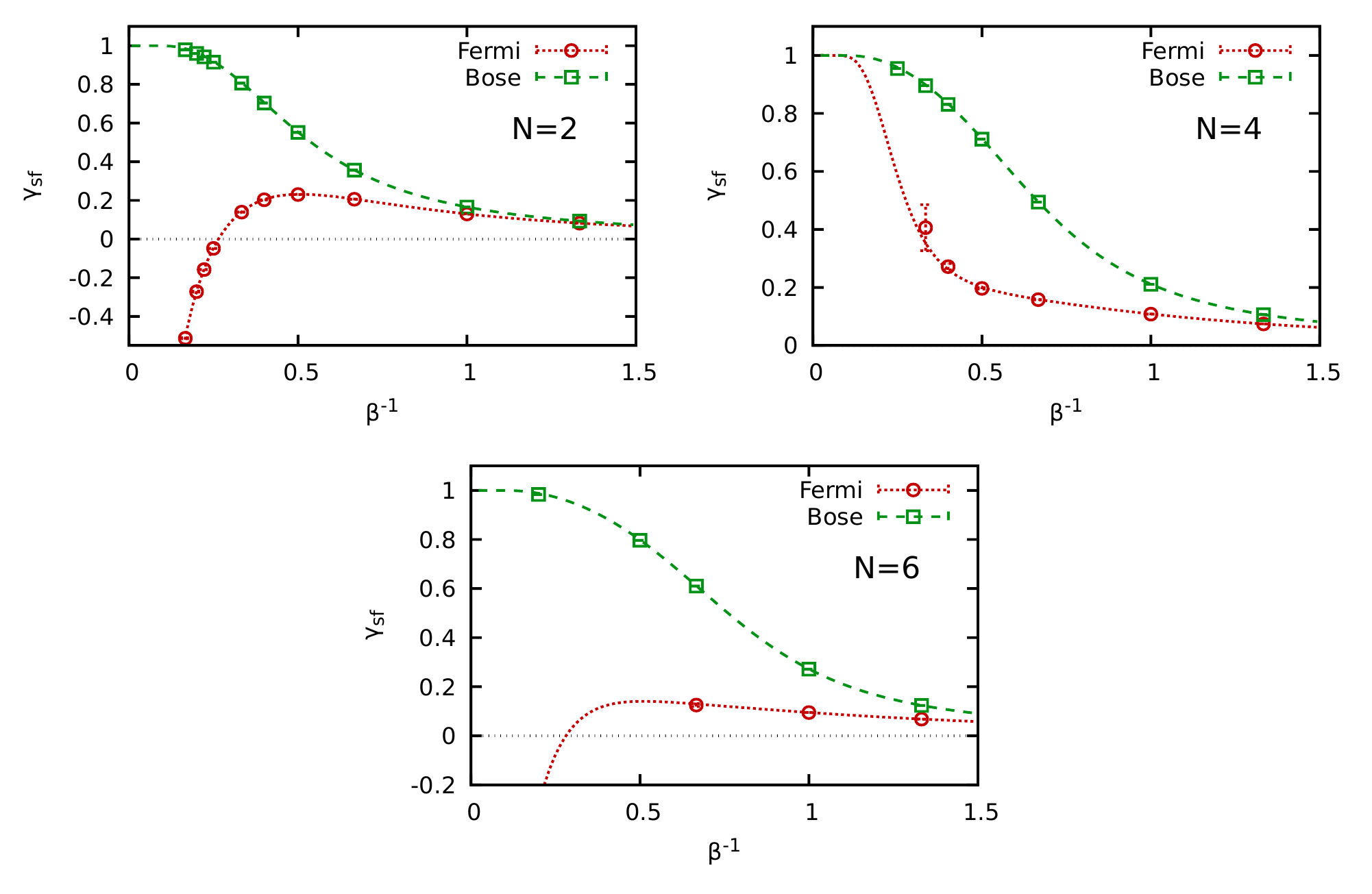}
\caption{\label{fig:ideal_SF_3D_N2}
Superfluid fraction $\gamma_\textnormal{sf}$ [cf.~Eq.~(\ref{eq:area_estimator})] for ideal fermions in $3D$. The symbols and lines show our numerical PIMC data and the exact result known from theory. Results for Bose- and Fermi-statistics are shown in green and red, respectively. Top left: $N=2$; top right: $N=4$; bottom: $N=6$.
}
\end{figure}

Let us start the discussion of our new simulation results with a brief study of the temperature dependence of the superfluid fraction of noninteracting bosons and fermions in a $3D$ harmonic trap. While similar analyses have been presented in 2D in Ref.~\cite{Dornheim_PRA_2020} and in 3D in Ref.~\cite{Blume_PRL_2014}, here we select somewhat different parameters. In addition, the availability of exact analytical results makes it a valuable test case to verify the implementation of our simulation scheme. Finally, the abnormal superfluid fraction of interacting electrons presented in a later section follows the same basic mechanism as the ideal case, which makes it a good starting point.

In Fig.~\ref{fig:ideal_SF_3D_N2}, we show results for $N=2$ (top left), $N=4$ (top right), and $N=6$ (bottom) ideal bosons (green) and fermions (red), with the symbols and lines showing PIMC data and the exact theoretical results. For all three system sizes, the superfluid fraction is small and almost equal for bosons and fermions for large temperatures $T=\beta^{-1}$, whereas differences emerge for smaller $T$. In fact, it always holds
\begin{eqnarray}
\lim_{T\to\infty}\gamma_\textnormal{sf} = 0\ ,
\end{eqnarray}
as the systems becomes purely classical in this limit, $I\to I_\textnormal{cl}$. Further, the first correction to $I_\textnormal{cl}$ is due to the finite extension $\lambda_\beta=\sqrt{2\pi\beta}$ of quantum particles, which is similar for both bosons and fermions.
With decreasing temperature, the paths of different particles in a PIMC simulation start to overlap, and the formation of permutation cycles, i.e., paths containing more than a single particle in them, becomes increasingly likely; see Ref.~\cite{dornheim_permutation_cycles} for a recent extensive discussion. In practice, this leads to an increased impact of quantum statistics, which explains the stark qualitative and quantitative differences between the red and green data sets. 

More specifically, bosons always attain a superfluid fraction of one in the limit of $T=0$ at these conditions, independent of the particle number $N$. Yet, we stress that this does certainly not mean that ideal bosons become an actual superfluid, as the Landau criterion requires a small value of the coupling parameter $\lambda$ in this case~\cite{landau}.
Fermions, on the other hand, exhibit a substantially more complicated behaviour. In particular, the superfluid fraction diverges towards negative infinity for both $N=2$ and $N=6$, and attains one for $N=4$, although for lower temperatures compares to bosons.
This remarkable effect has been explained by Yan and Blume~\cite{Blume_PRL_2014} in terms of the structure of the density matrix. Specifically, it is the presence of an energetically low-lying state with a finite circulation for some $N$ that leads to a diverging moment of inertia and to the observed behaviour in $\gamma_\textnormal{sf}$.

Finally, we stress the excellent agreement between theory and PIMC data for both bosons and fermions everywhere, which is a strong verification of our implementation and analysis. In addition, we note that simulations are restricted to above a minimum temperature in the case of fermions due to the fermion sign problem (see Sec.~\ref{sec:PIMC} above), as can be seen most clearly for the leftmost red data point for $N=4$. In contrast, there is no such issue for bosons, where simulations are possible even at extremely low temperatures.

\subsection{Abnormal superfluid fraction of electrons: dependence on temperature\label{sec:temperature}}

\begin{figure}\centering
\includegraphics[width=0.51\textwidth]{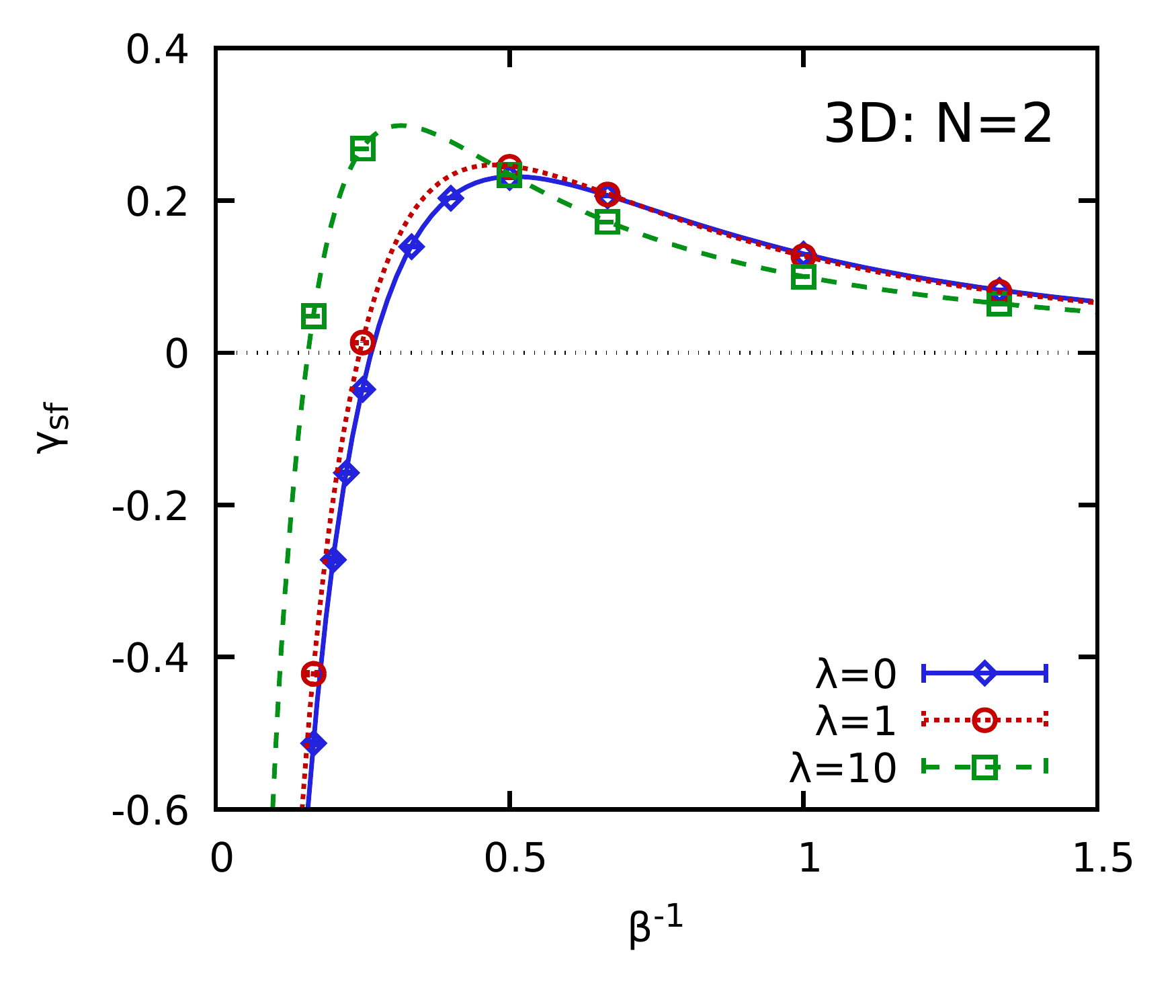}
\caption{\label{fig:two_particle}
Exact results for the temperature-dependence of the superfluid fraction $\gamma_\textnormal{sf}$ for $N=2$ particles in a $3D$ harmonic trap. Solid blue: $\lambda=0$ (ideal); dotted red: $\lambda=1$; dashed green: $\lambda=10$. The corresponding symbols show PIMC data for the same parameters.
}
\end{figure}  

Let us begin our investigation of interacting electrons by studying the two-particle Hamiltonian. The relative and center of mass coordinates are separable; the center-of-mass energy spectrum is solved analytically and the relative energy spectrum is  obtained by the finite-element method. Eigen energies subject to a relative energy cutoff of around $100\hbar\Omega$ are computed. We calculate the actual moment of inertia $I$ through $I=\hbar^{2}\left\langle M^{2}\right\rangle_{\mathrm{th}} /\left(k_{B} T\right)$, where $M$ is the angular momentum projection quantum number and $\left\langle \right\rangle$ indicates the thermal average~\cite{Blume_PRL_2014}. The classical moment of inertia $I_{\mathrm{cl}}$ is obtained from the thermal average $\left\langle \sum_{k=1}^{N}r_{k,\perp}^2\right\rangle$. Here $r_{k,\perp}^2$ is the square of distance of the $k$-th particle to the $z$ axis. Due to spherical symmetry, it is related to the trap energy $\left\langle \sum_{k=1}^{N}r_{k,\perp}^2\right\rangle_{\mathrm{th}}=\frac{4}{3}\left\langle \sum_{k=1}^{N}\frac{1}{2}r_{k,\perp}^2\right\rangle_{\mathrm{th}}$ where the trap energy is then evaluated by converting it to the thermal average of the derivative of the energy with respect to the trap depth through the Hellmann–Feynman theorem.
The solid blue, dotted red, and dashed green lines in Fig.~\ref{fig:two_particle} show the numerical results for $\lambda=0$, $\lambda=1$, and $\lambda=10$.
First and foremost, we note that these results are in excellent agreement to our new PIMC results, which are depicted by the corresponding symbols. Secondly, we find that the coupling strength has a comparatively small impact on the behaviour of $\gamma_\textnormal{sf}$ even for $\lambda=10$; indeed, the results for moderate coupling ($\lambda=1$) can hardly be distinguished from the ideal curve.
This is a strong indication that the response of electrons in a quantum dot to an external rotation is largely determined by the properties of the corresponding noninteracting system.

\begin{figure}\centering
\includegraphics[width=0.99\textwidth]{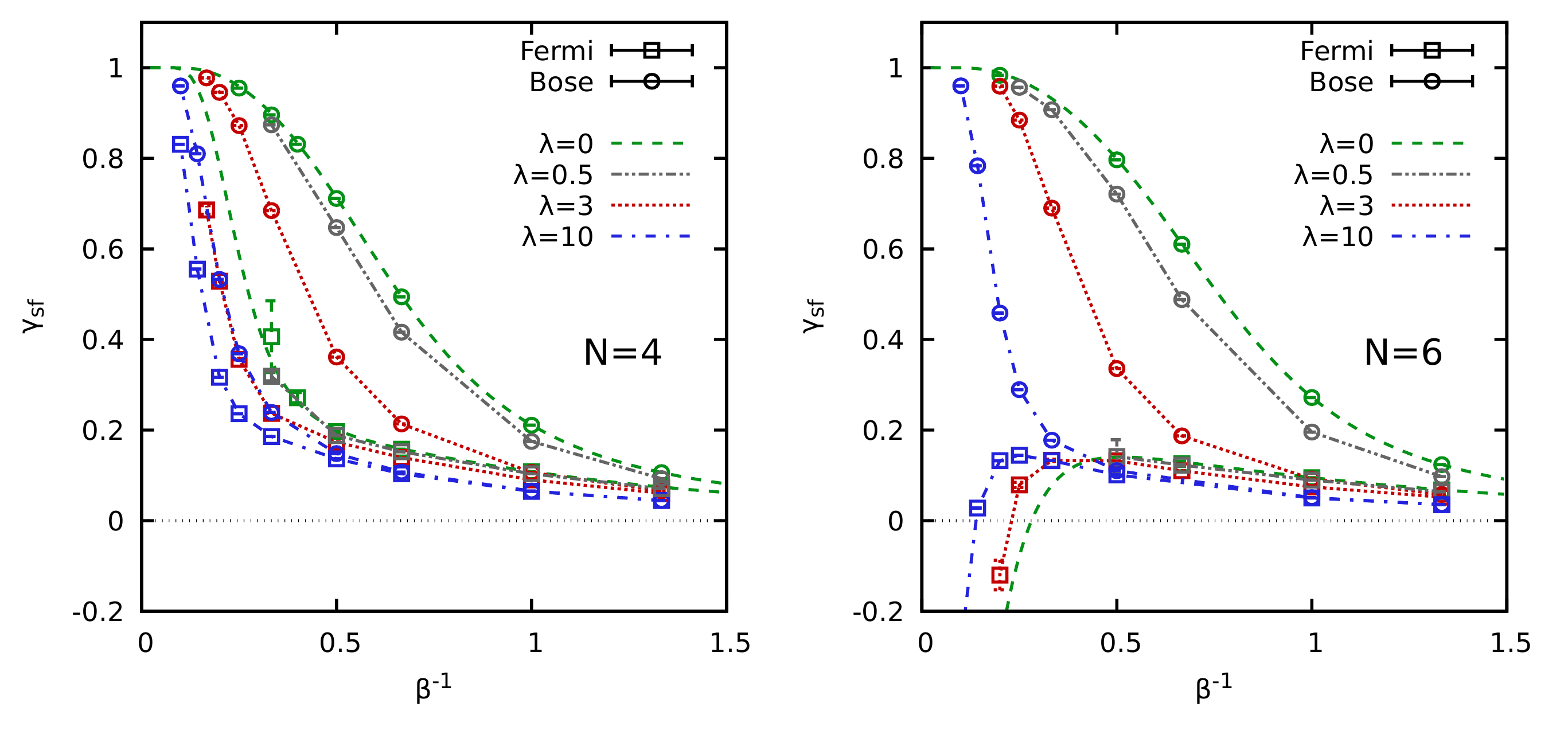}
\caption{\label{fig:3D_Superfluid}
Temperature dependence of the superfluid fraction $\gamma_\textnormal{sf}$ for $N=4$ (left) and $N=6$ (right) electrons in a $3D$ harmonic trap. The dashed green, dash-double-dotted grey, dotted red, and dash-dotted blue curves correspond to $\lambda=0$ (ideal), $\lambda=0.5$, $\lambda=3$, and $\lambda=10$, respectively. In addition, results for bosons and fermions are depicted as circles and squares.
}
\end{figure}

Let us next consider our new PIMC data for the temperature dependence of  $\gamma_\textnormal{sf}$ for larger systems that is shown in Fig.~\ref{fig:3D_Superfluid}. The left panel shows results for $N=4$ particles, and the squares and circles have been obtained for Fermi- and Bose-statistics. The green curves show results for the ideal case ($\lambda=0$) already discussed in Fig.~\ref{fig:ideal_SF_3D_N2} above and have been included as a reference. 
The dash-double-dotted grey curves have been obtained for a small yet finite value of the coupling parameter, $\lambda=0.5$, and closely follow the ideal case. In particular, the fermionic results cannot be distinguished from the noninteracting curve with the bare eye, whereas the gap for Bose-statistics is significant.

Going to moderate coupling strength, $\lambda=3$ (dashed red), shifts the curves substantially to the left compared to smaller $\lambda$. Heuristically, this can be understood as follows: due to the stronger Coulomb repulsion, the particles are more clearly separated. Consequently, it takes lower temperatures until paths of different particles overlap, which is the driving mechanism of a large superfluid fraction~\cite{dornheim_superfluid,Filinov_PRB_2008}.
Again, we find that the effect of the Coulomb repulsion is more pronounced in the case of bosons compared to fermions.
The reason for this effect has been reported by Dornheim~\cite{Dornheim_PRA_2020} for the case of dipole--dipole interaction: for fermions, weak interactions are effectively masked by the Pauli exclusion principle which separates even ideal fermions from each other; in contrast, bosons tend to cluster next to each other, and are thus more strongly affected by the Coulomb repulsion. A more detailed analysis of this effect is presented in Sec.~\ref{sec:coupling} below.

Finally, the blue curves correspond to relatively strong coupling, $\lambda=10$. Interestingly, the curves for the two different types of quantum statistics are quite close to each other in this case, and both approach unity for similar temperatures.

Let us next consider the same property, but for $N=6$ shown in the right panel of Fig.~\ref{fig:3D_Superfluid}. For bosons, we find a quite similar picture compared to the left panel. Evidently, the particular particle number does not substantially shape the qualitative physical behaviour at these conditions.
In stark contrast, all fermionic curves attain a maximum in $\gamma_\textnormal{sf}$ in the range $0.5\leq \beta \leq  5$ until they diverge towards negative infinity for $T\to0$. Still, the impact of the coupling parameter $\lambda$ is only quantitative, and the main feature is already present in the ideal case. Furthermore, we again find that fermions are less affected by the Coulomb repulsion compared to bosons.

\begin{figure}\centering
\includegraphics[width=0.99\textwidth]{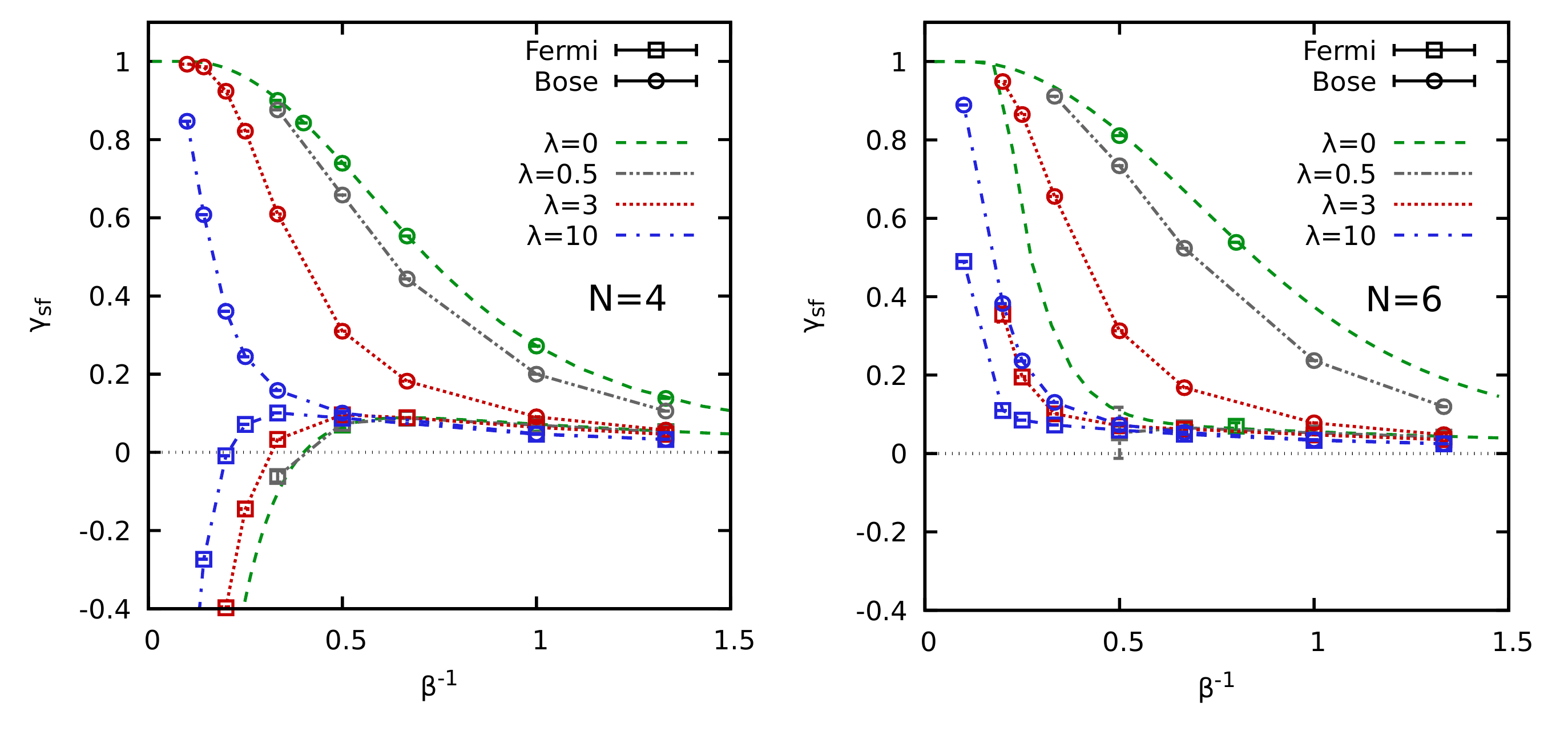}
\caption{\label{fig:2D_Superfluid}
Temperature dependence of the superfluid fraction $\gamma_\textnormal{sf}$ for $N=4$ (left) and $N=6$ (right) electrons in a $2D$ harmonic trap. The dashed green, dash-double-dotted grey, dotted red, and dash-dotted blue curves correspond to $\lambda=0$ (ideal), $\lambda=0.5$, $\lambda=3$, and $\lambda=10$, respectively. In addition, results for bosons and fermions are depicted as circles and squares. The data for the ideal system have partly been taken from Ref.~\cite{Dornheim_PRA_2020}.
}
\end{figure}

In Fig.~\ref{fig:2D_Superfluid}, we present a similar analysis, but for a purely $2D$ system. The results are very similar to the $3D$ case, but the $T=0$ limit is flipped in the case of fermions: for $N=4$, $\gamma_\textnormal{sf}$ diverges towards negative infinity, and for $N=6$, it attains unity just as in the Bose-case. 
Still, the phenomenological reason is the same for both $2D$ and $3D$: it is the presence of a state with finite circulation~\cite{Blume_PRL_2014} in the respective cases that causes the negative divergence of the superfluid fraction.

\begin{figure}\centering
\includegraphics[width=0.99\textwidth]{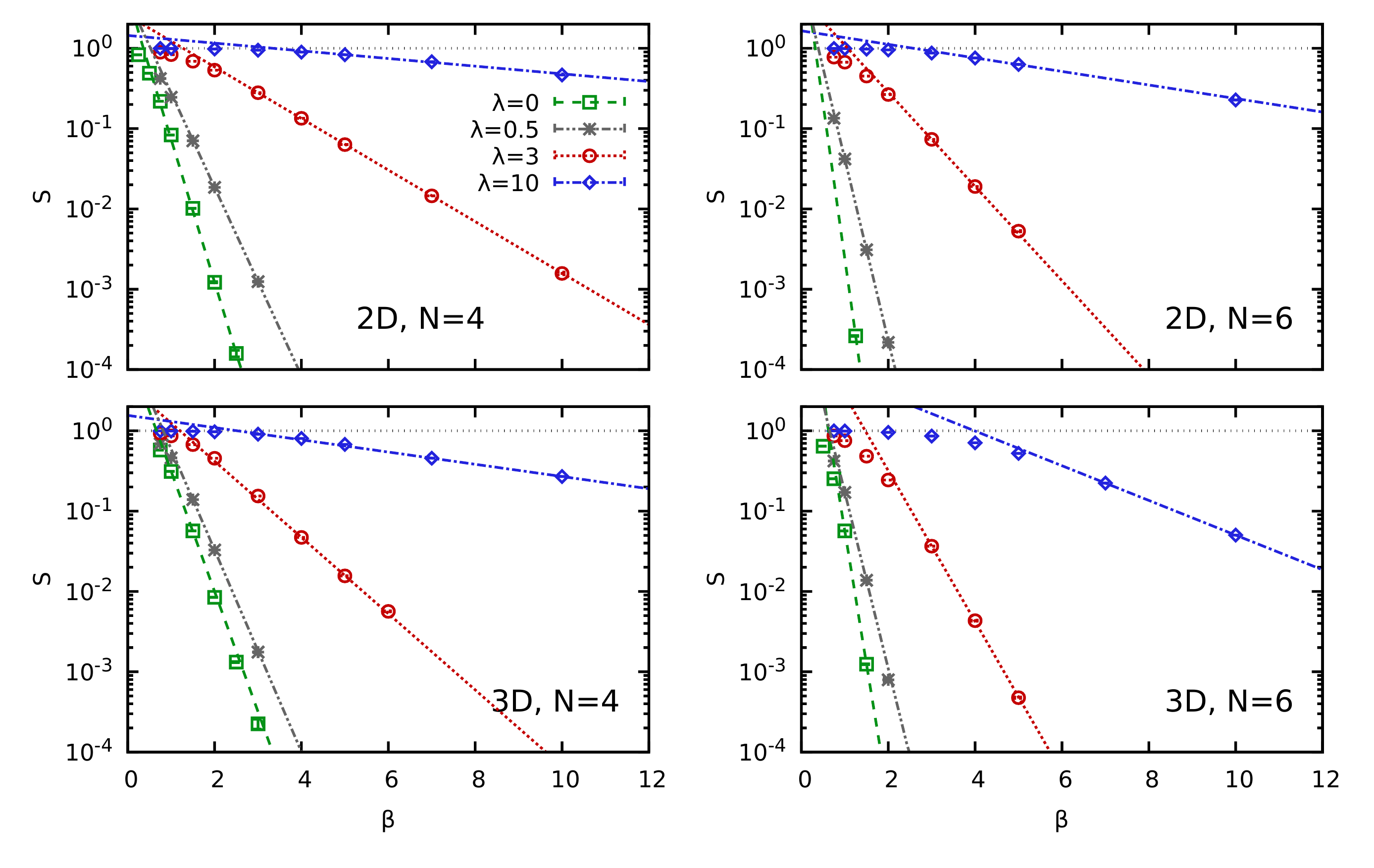}
\caption{\label{fig:Sign_panel}
Average sign from the PIMC simulations from Figs.~\ref{fig:2D_Superfluid} (top row) and \ref{fig:3D_Superfluid} (bottom row). The straight lines correspond to exponential fits according to Eq.~(\ref{eq:sign_fit}).
}
\end{figure}

Let us briefly postpone our analysis of the connection of the abnormal superfluid fraction of fermions to the structural properties of the system to touch upon the manifestation of the fermion sign problem.
In Fig.~\ref{fig:Sign_panel}, we show the PIMC expectation values for the denominator in Eq.~(\ref{eq:ratio}), i.e., the average sign. The top and bottom rows have been obtained for $2D$ and $3D$ systems, and the left and right columns correspond to $N=4$ and $N=6$ spin-polarized electrons. Furthermore, the different data points distinguish data for various values of the coupling parameter $\lambda$. First and foremost, we note that all depicted curves exhibit the same qualitative behaviour: for small $\beta$ (large temperatures), the extension of the paths of different particles is substantially smaller than the average inter-particle separation. Consequently, permutation cycles only rarely appear in the PIMC simulation, most configuration weights are positive, and the average sign is large. With decreasing temperature, the thermal wavelength increases as $\lambda_\beta\sim\sqrt{\beta}$ and, eventually, the paths of all particles in the system overlap. This makes the formation of permutation cycles significantly more likely, configurations with positive and negative weights appear with a similar frequency in the PIMC simulation, and the average sign drops.
The straight lines in Fig.~\ref{fig:Sign_panel} have been obtained from exponential fits according to
\begin{eqnarray}\label{eq:sign_fit}
S(\beta) = a e^{-b\beta}\ 
\end{eqnarray}
above an empirical minimum value of $\beta$, with $a$ and $b$ being the free parameters. Evidently, the fits are in excellent agreement to the PIMC data, which confirms the exponential nature of the fermion sign problem reported in earlier studies~\cite{dornheim_sign_problem}.

\begin{figure}\centering
\includegraphics[width=0.99\textwidth]{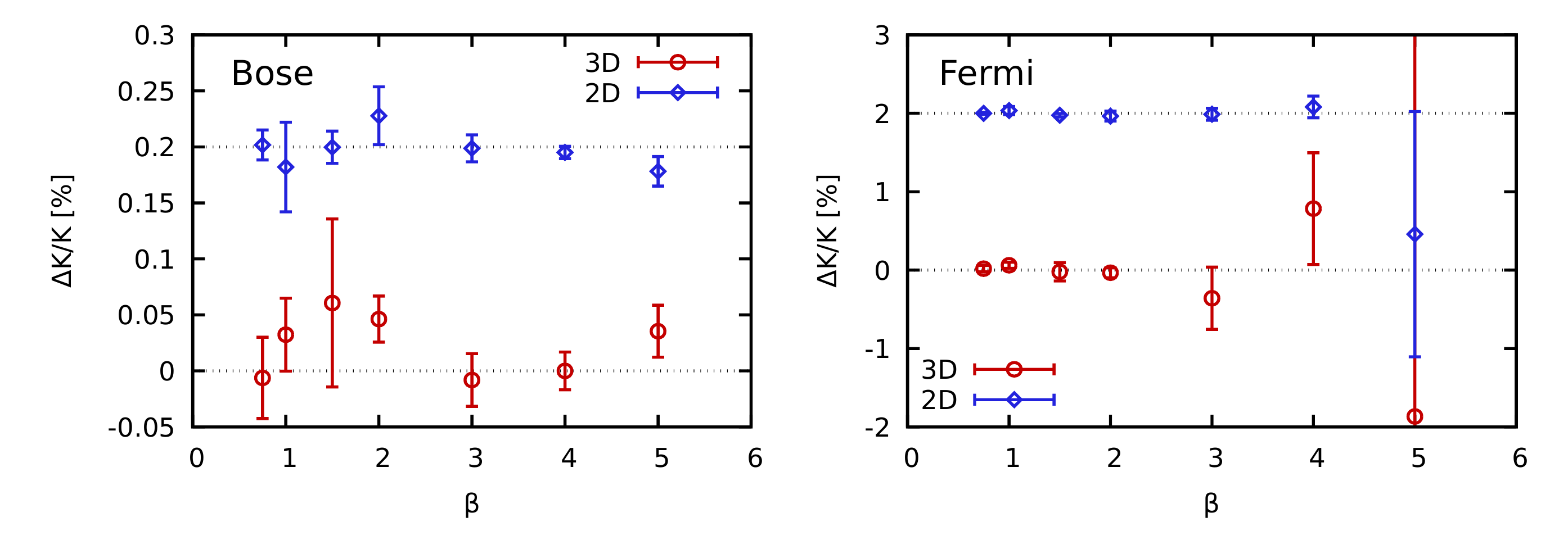}
\caption{\label{fig:virial}
Relative difference in the kinetic energy $K$ (in per cent) between the direct PIMC estimator and the virial theorem [Eq.~(\ref{eq:K_vir})] for $N=6$ electrons at $\lambda=3$ as a function of the inverse temperature $\beta$. The red circles and blue diamonds correspond to the $3D$ and $2D$ case, and the latter have been shifted upwards (horizontal dotted grey line) for better visibility. The left and right panels have been obtained for Bose- and Fermi-statistics.
}
\end{figure}  

A related interesting point is the PIMC estimation of the corresponding energies. More specifically, we consider the well-known 
virial theorem~\cite{greiner1995thermodynamics}, which gives a relation between the kinetic ($K$), interaction ($W$), and external potential ($v_\textnormal{ext}$) contributions to the total energy,
\begin{eqnarray}\label{eq:K_vir}
K = V_\textnormal{ext} - \frac{W }{2} \quad .
\end{eqnarray}
In practice, we directly estimate $K$ in our PIMC simulation using the standard thermodynamic estimator (see, e.g., Ref.~\cite{cep}) and compare it to the right-hand side of Eq.~(\ref{eq:K_vir}). The results are depicted in Fig.~\ref{fig:virial}, where we show the relative deviation between the two estimations of $K$ in per cent. We note that the virial theorem as it is given in Eq.~(\ref{eq:K_vir}) is a general property of the Hamiltonian and holds independent of the particular type of quantum statistics. The left panel has been obtained for Bose-statistics, and the red circles and blue diamonds show results for the $3D$ and $2D$ case, where the latter have been shifted by $0.2\%$ for better visibility.
Evidently, all data points fluctuate around zero deviation within the given level of statistical uncertainty. Further, we find that the error bars are only weakly dependent on $\beta$, and are of the order of $\Delta K/K\sim 0.01\%$.

The right panels shows results from the same simulation computed for Fermi-statistics by evaluating Eq.~(\ref{eq:ratio}). Again, we find that the data points fluctuate around zero deviation between the two different estimations of $K$, which is a strong verification of our simulation scheme and its implementation. Yet, we find a drastic increase of the error bars with $\beta$ as predicted by Eq.~(\ref{eq:FSP}). This nicely illustrates the relative difficulty of fermionic PIMC simulations compared to the Bose-case, where this problem is absent. In particular, simulations of as few as four electrons become unfeasible even on modern supercomputers when the temperature is decreased.

\begin{figure}\centering
\includegraphics[width=0.99\textwidth]{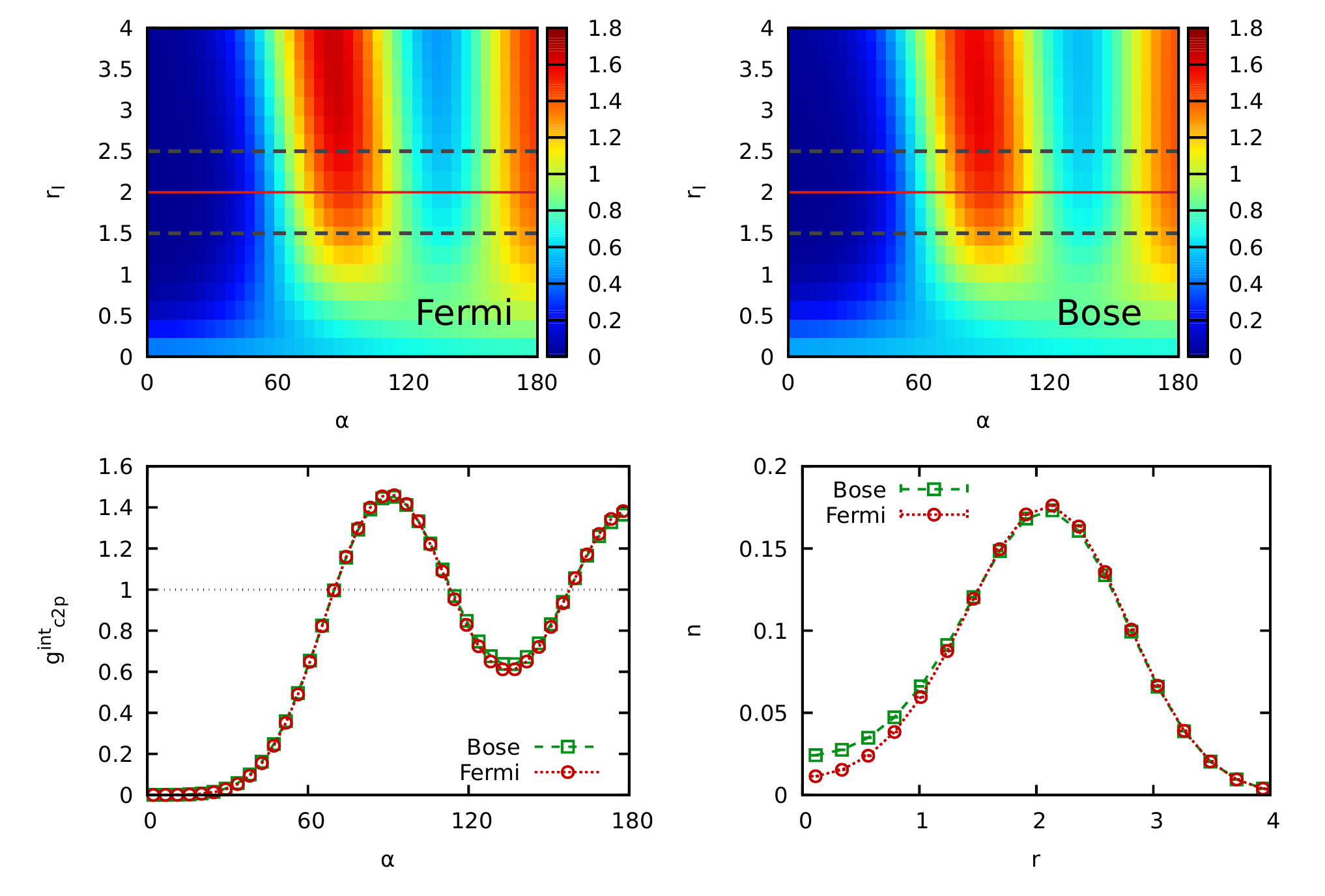}
\caption{\label{fig:c2p_N4_lambda10}
Top row: integrated C2P for $N=4$ electrons at $\lambda=10$ and $\beta=10$ for the range $1.5\leq r_{II} \leq 2.5$ (horizontal dashed grey lines) for fermions (left) and bosons (right). Bottom left: Scan-line of the integrated C2P at $r_I=2$. Bottom right: radial density profile $n(r)$.
}
\end{figure}

Let us next consider the structural properties of electrons in a $2D$ quantum dot, and their relation to the abnormal superfluid fraction reported above. To this end, we estimate the integrated C2P defined in Eq.~(\ref{eq:integrated_c2p}) in Sec.~\ref{sec:C2P}. The results are shown in Fig.~\ref{fig:c2p_N4_lambda10} for $N=4$ at $\lambda=10$ and $\beta=10$. The top row shows the integrated C2P itself in the $\alpha$-$r_I$-plane, with the reference particle being located somewhere in the range $1.5\leq r_{II} \leq 2.5$ (horizontal dashed grey lines), i.e., around the maximum of the radial density (see the bottom right panel of Fig.~\ref{fig:c2p_N4_lambda10}). Further, the left and right panels show results for fermions and bosons, respectively.
Naturally, the C2P is fully symmetric in $\alpha$ around $180$ degrees and we restrict ourselves to this half-range. Most importantly, we find that the type of quantum statistics has only very mild effects on the structural properties on the system at these conditions. This is a direct consequence of the strong coupling strength, which effectively separates the paths of individual particles from each other in the PIMC simulation and suppresses the formation of permutation cycles~\cite{dornheim_permutation_cycles}.
For small angular separations $\alpha$, the integrated C2P exhibits a pronounced exchange--correlation hole, followed by a maximum at $\alpha=90$, a subsequent minimum which is substantially less pronounced than the exchange--correlation hole itself, and a second maximum at the opposite end of the system, i.e., around $\alpha=180$. This can be seen particularly well in the bottom left panel of Fig.~\ref{fig:c2p_N4_lambda10}, where we show a scan-line over the C2P at $r_I=2$, see the solid red line in the top row.
The red circles and green squares show results for fermions and bosons, but the deviations between the two data sets can barely be resolved with the naked eye.
From a physical perspective, this indicates that the most common configuration of particles is given by a single ring around the center of the trap containing all four particles. This is further substantiated by the radial density profile $n(r)$ shown in the bottom right panel of Fig.~\ref{fig:c2p_N4_lambda10}. Here the main impact of quantum statistics can be found around $r=0$, where the probability to find a fermionic particle is reduced compared to the case of bosons.

Let us now connect these findings regarding the structural properties of the system to the abnormal superfluid fraction investigated above. In particular, it can be seen in Fig.~\ref{fig:2D_Superfluid} that $\gamma_\textnormal{sf}$ strongly depends on the type of quantum statistics at these conditions. Remarkably, this is not reflected in the structural properties of the system, which, indeed, can hardly be distinguished at all.

\subsection{Particle number dependence\label{sec:N}}

A further interesting topic of investigation is the dependence of the properties of small electronic systems on the number of particles. In this context, we mention that the properties of confined few-particle systems often sensitively depend on $N$, which can be seen particularly well in the investigation of addition energies~\cite{Addition1,Addition2,Addition3,Addition4}. Furthermore, Filinov \emph{et al.}~\cite{Filinov_PRB_2008} have found that the nonclassical rotational inertia of strongly coupled charged bosons exhibits certain "magic numbers", where the superfluid fraction is most pronounced.
In this work, we provide a similar analysis for the superfluid fraction $\gamma_\textnormal{sf}$ of electrons in quantum dots.

\begin{figure}\centering
\includegraphics[width=0.99\textwidth]{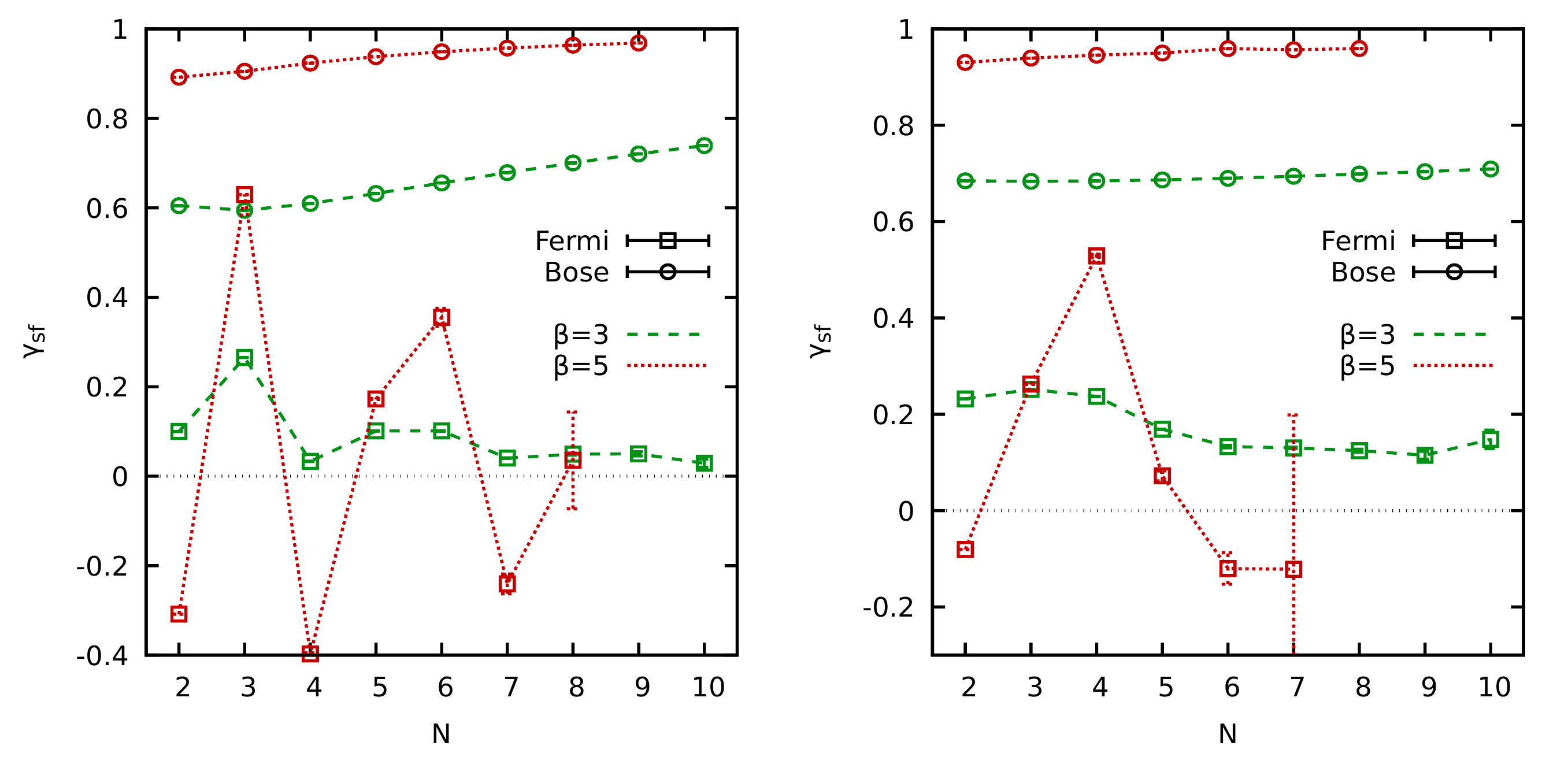}
\caption{\label{fig:Superfluid_N}
Superfluid fraction in $2D$ (left) and $3D$ (right). Shown is the particle number dependence for $\lambda=3$ with the squares and circles corresponding to Fermi- and Bose-statistics, and the dashed green and dotted red lines showing results for $\beta=3$ and $\beta=5$, respectively.
}
\end{figure}

The results are shown in Fig.~\ref{fig:Superfluid_N}, where we show the $N$-dependence of $\gamma_\textnormal{sf}$ both in $2D$ (left panel) and $3D$ (right panel) at moderate coupling, $\lambda=3$. The circles and squares show results for Bose- and Fermi-statistics, and the red and green curves have been obtained for $\beta=5$ and $\beta=3$. For all depicted cases, the bosonic curves are quite smooth and exhibit a monotonous increase (the small exception being $N=3$ for $\beta=3$ in $2D$) in the superfluid fraction with $N$. Heuristically, this can be understood as follows: in the harmonic trap, the average density increases with $N$. This, in turn, leads to an increased value of the degeneracy parameter $\chi=\lambda_\beta/\overline{r}$, where $\overline{r}$ is the average inter-particle distance~\cite{dornheim_superfluid}, and thus to an increasing value of $\gamma_\textnormal{sf}$.

For fermions, on the other hand, the situation is significantly more complicated and interesting. In the $2D$-case, the superfluid fraction is rapidly oscillating with $N$, which is especially pronounced for $\beta=5$. In particular, the negative and positive values of $\gamma_\textnormal{sf}$ can be traced back to the ideal case, see Ref.~\cite{Dornheim_PRA_2020} for such data in two dimensions. A second trend, which can be seen most clearly for $\beta=3$, is given by the overall decrease of the superfluid fraction with $N$. This, too, is expected by simple physical intuition. An increase in $N$ leads to a transition from a finite to a bulk system. For bosons, the crossover from a normal to a superfluid system then attains the character of a true phase transition under the right conditions~\cite{dornheim_superfluid}. In contrast, a single species of fermions such as spin-polarized electrons cannot exhibit the required off-diagonal long-range order~\cite{Shi_PRB_2005}, and a corresponding bulk system will thus not be superfluid. This is reflected in the observed decrease in $\gamma_\textnormal{sf}$ with $N$. For completeness, we mention that superfluidity in Fermi-systems is still possible, but requires a pairing mechanism such as the interaction with phonons leading to superconductivity~\cite{Bardeen_PhysRev_1957}.

We find similar trends in the case of fermions for $3D$ systems, although both the oscillations with $N$ and the decrease with increasing system size are somewhat less pronounced.

\begin{figure}\centering
\includegraphics[width=0.5\textwidth]{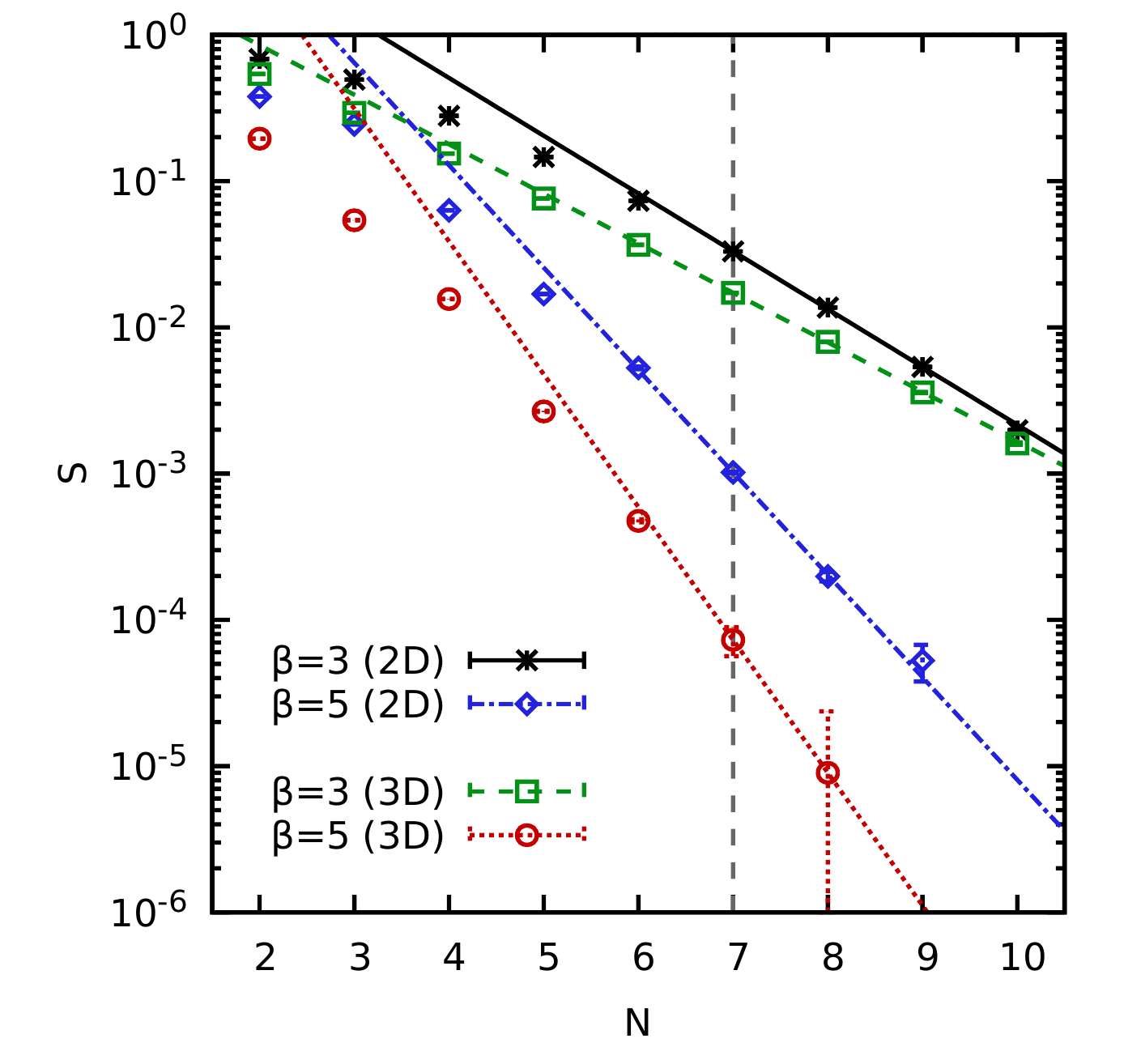}
\caption{\label{fig:Sign_N}
Particle number dependence of PIMC results for the average sign $S$ for $\lambda=3$. The black stars (green squares) and blue diamonds (red circles) have been obtained for $\beta=3$ and $\beta=5$, respectively, for spin-polarized electrons in a $2D$ ($3D$) quantum dot. The lines correspond to exponential fits for $N\geq7$ (vertical dashed grey line), cf.~Eq.~(\ref{eq:sign_fit_beta}). 
}
\end{figure}

Let us conclude this investigation of the $N$-dependence by briefly revisiting the fermion sign problem. To this end, we show the average sign obtained for the simulations shown in Fig.~\ref{fig:Superfluid_N} in Fig.~\ref{fig:Sign_N}. The points show the PIMC data, and the straight lines correspond to exponential fits of the form
\begin{eqnarray}\label{eq:sign_fit_beta}
S(N) = a e^{-bN}\ ,
\end{eqnarray}
and have been obtained by introducing an empirical minimum value of $N=7$ (vertical dashed grey line). Strictly speaking, Eq.~(\ref{eq:sign_fit_beta}) is only expected to hold for a bulk system such as the uniform electron gas~\cite{dornheim_sign_problem,review}, whereas the dependence of a finite, trapped system is expected to be more complicated. Still, we do find at least a qualitative agreement with the exponential decay. The particularly small value of $S$ for the $3D$ case at $\beta=5$ is directly reflected by the large error bars in $\gamma_\textnormal{sf}$ observed in Fig.~\ref{fig:Superfluid_N}. This again strongly illustrates the severe limitations of fermionic PIMC simulations due to the notorious sign problem.

\subsection{Dependence on coupling strength\label{sec:coupling}}

\begin{figure}\centering
\includegraphics[width=0.99\textwidth]{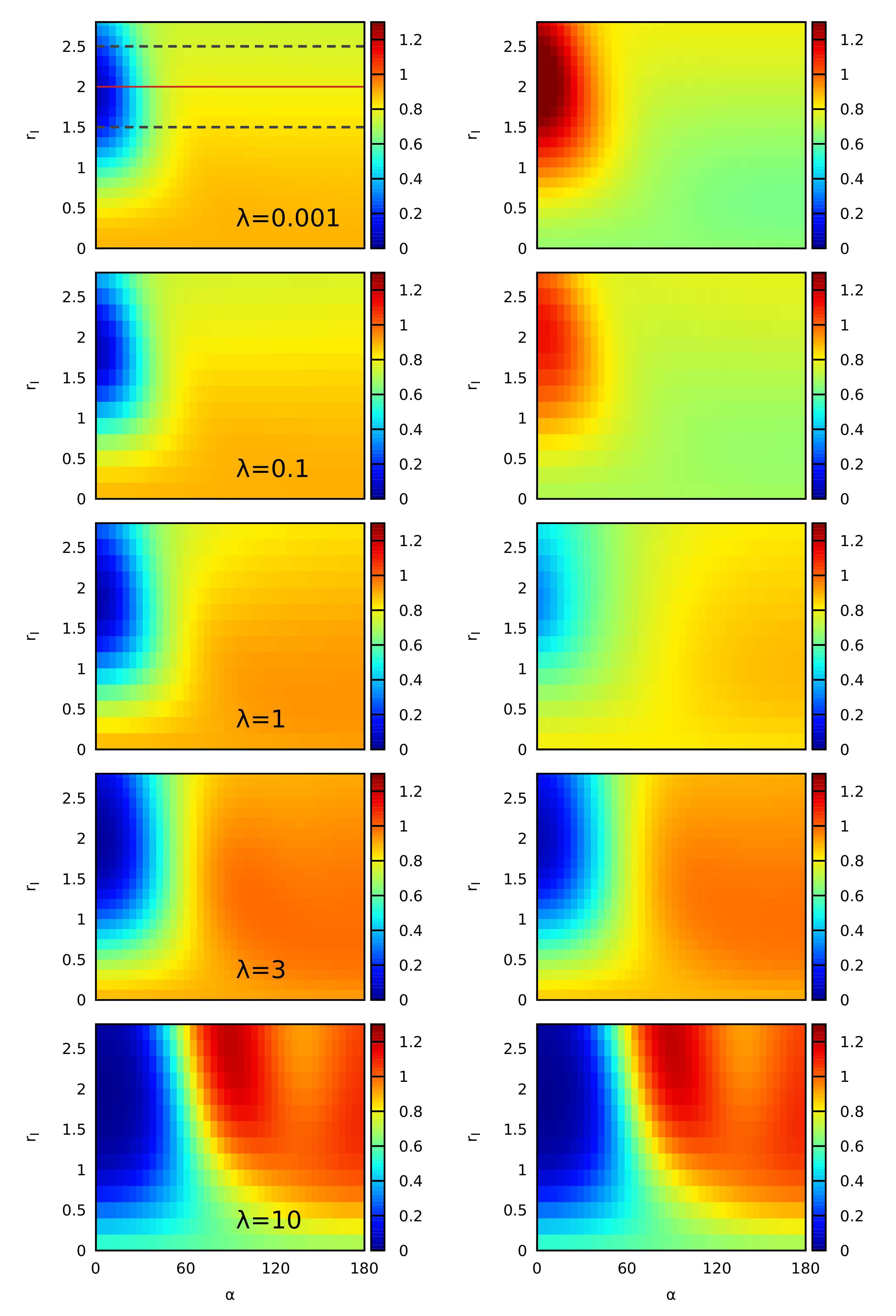}
\caption{\label{fig:c2p_N4_lambda}
Integrated C2P for $N=4$ spin-polarized electrons in a harmonic confinement for $\beta=1$ in $2D$ for Fermi- (left) and Bose-statistics (right).
}
\end{figure}

A highly interesting question is the precise nature of the intricate interplay of the Coulomb repulsion with the effect of quantum statistics. In this regard, our computationally expensive PIMC simulations are uniquely suited to provide reliable answers as no approximations are involved.  
In Fig.~\ref{fig:c2p_N4_lambda}, we show results for the integrated C2P [cf.~Eq.~(\ref{eq:integrated_c2p}) above] for $N=4$ spin-polarized electrons in a $2D$ harmonic trap at a moderate temperature, $\beta=1$. The left and right columns have been obtained for Fermi- and Bose-statistics, and the reference particle is always located in the interval $1.5\leq r_{II}\leq 2.5$ (horizontal dashed grey lines in the top left panel).

The top row corresponds to a very weakly coupled system ($\lambda=10^{-3}$), where the Coulomb repulsion has a negligible effect. For fermions, the Pauli exclusion principle nevertheless results in a pronounced exchange--correlation hole around $\alpha=0$, whereas the rest of the system appears to be relatively featureless. In stark contrast, a corresponding Bose-system at the same conditions exhibits the opposite behaviour, as bosons tend to cluster around each other. 
Increasing the coupling strength to $\lambda=0.1$ leaves the Fermi-system unchanged, whereas the positive \emph{exchange--correlation hill} of the bosons is substantially decreased. From a physical perspective, this means that bosons are strongly influenced even by weak repulsive interactions, whereas the latter are effectively masked by the Pauli principle in the case of fermions. 
For completeness, we mention that the same point has recently been reported by Dornheim~\cite{Dornheim_PRA_2020} in the case of dipole--dipole repulsion.
The same trend has also been noted in our discussion of Figs.~\ref{fig:2D_Superfluid} and \ref{fig:3D_Superfluid} above, where the superfluid fraction $\gamma_\textnormal{sf}$ was substantially more affected by the coupling parameter $\lambda$ for bosons compared to fermions.

Going back to the integrated C2P shown in Fig.~\ref{fig:c2p_N4_lambda}, a further increase of the coupling strength to $\lambda=1$ (middle row) again only leads to very mild changes for Fermi-statistics, whereas there appears a qualitative change for bosons: the exchange--correlation \emph{hill} is changed to a \emph{hole}, although it is still significantly less pronounced compared to the left panel.
At $\lambda=3$, i.e., at moderate coupling, the importance of quantum statistics is drastically reduced and both panels give a nearly identical picture. Finally, the strong Coulomb repulsion at $\lambda=10$ effectively separates individual particles, and no differences between bosons and fermions can be seen with the naked eye.

\begin{figure}\centering
\includegraphics[width=0.5\textwidth]{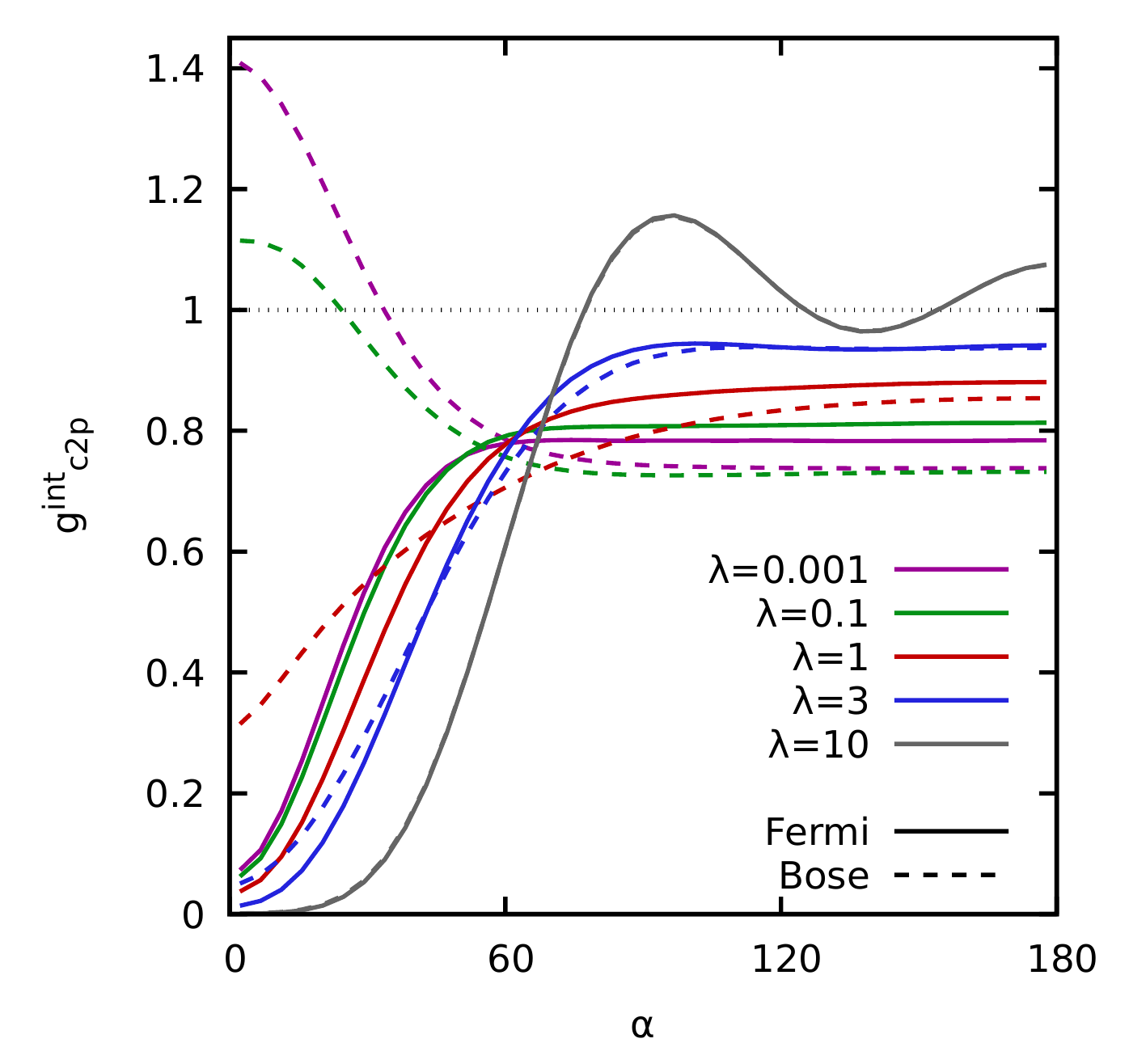}
\caption{\label{fig:density_N4_lambda}
Scan-line for Fig.~\ref{fig:c2p_N4_lambda}.
}
\end{figure}

Let us next consider Fig.~\ref{fig:density_N4_lambda}, where we show a scan-line of the integrated C2P at $r_I=2$ (solid red line in the top left panel of Fig.~\ref{fig:c2p_N4_lambda}). The solid and dashed lines distinguish fermions and bosons, and the different colours correspond to values of the coupling parameter $\lambda$. The arguably most striking feature of this plot is the \emph{exchange--correlation hill} for bosons at $\lambda\leq0.1$. In addition, it clearly illustrates the small impact of the coupling strength on fermions for the same $\lambda$. For $\lambda=3$, there is still a distinct impact of quantum statistics, whereas no such difference can be resolved for $\lambda=10$. The latter case also constitutes the only curve with a second maximum around $\alpha=180$.
Let us close this discussion of the integrated C2P with a practical remark. Obviously, the probability to find two identical fermions at the exact same position must always be zero, $\rho_2(r,r,0)=0$. Yet, by integrating over a finite range of the reference particle $r_{II,\textnormal{min}}\leq r_{II} \leq r_{II,\textnormal{max}}$, the integrated C2P  can be nonzero for all values of $r_I$ even for $\alpha=0$.

\section{Summary and Outlook\label{sec:summary}}

In summary, we have presented extensive new \emph{ab initio} path-integral Monte Carlo simulation results for electrons in $2D$ and $3D$ quantum dots. In particular, we have investigated the nonclassical rotational inertia of such systems and have found a divergent, negative abnormal superfluid fraction which manifests under certain conditions due to the interplay of Fermi-statistics and the harmonic confinement~\cite{Blume_PRL_2014}. In addition, we have thoroughly investigated the structural properties of such systems by computing the center-two particle correlation function suggested by Thomsen and Bonitz~\cite{Thomsen_PRE_2015}. The comparison of results for Bose- and Fermi-statistics has revealed that the onset of the abnormal superfluid fraction is hardly connected to the static structure, which can be nearly independent of the type of quantum statistics, whereas the respective superfluid fractions diverge. 
Instead, it is a direct consequence of the structure of the density matrix with respect to the angular momentum.
In addition, we have substantiated earlier findings for harmonically confined quantum dipole systems~\cite{Dornheim_PRA_2020}, where it has been found that the repulsion between two fermions is effectively masked by the Pauli exclusion principle for weak too moderate values of the coupling parameter $\lambda$. In stark contrast, bosons sensitively react even to small changes in the coupling strength.

Future extensions of our work might include the investigation of spin-unpolarized systems, which would allow for the study of a spin-resolved C2P. In particular, it is clear that the observed effective screening of the pair interaction by the Pauli principle will not apply to fermions of a different spin-orientation, which will lead to an interesting mixture of different effects. In addition, we note that the central computational bottleneck of the present study is given by the fermion sign problem, which precludes the simulation of larger systems and lower temperatures. In this regard, we mention the remarkable recent progress in the fermionic quantum Monte Carlo simulation at finite temperature in the context of warm dense matter research~\cite{Malone_JCP_2015,Malone_PRL_2016,Dornheim_NJP_2015,Dornheim_PRB_2016,dornheim_prl,groth_prl,Brown_ethan,Rubenstein_auxiliary_finite_T,Clark_PRB_2017,Blunt_PRB_2014,yilmaz2020restricted,doi:10.1063/5.0041378,PhysRevE.103.053204}. Specifically, the adaption of these novel methods to the simulation of electrons in quantum dots seems promising and rewarding.
Finally, we mention that the C2P that has been used in the present work allows to reliably describe crossovers and melting phenomena in finite systems~\cite{Thomsen_PRE_2015}. This is in stark contrast to previously employed Lindemann-type approaches~\cite{PhysRevLett.100.113401} that rely on the underlying Monte Carlo sampling scheme. Thus, the present set-up can be straightforwardly used to study the impact of quantum effects onto the crystallization of trapped few-body systems.

\section*{Acknowledgments}

  This work was partially funded by the Center for Advanced Systems Understanding (CASUS) which is financed by Germany’s Federal Ministry of Education and Research (BMBF) and by the Saxon Ministry for Science, Culture and Tourism (SMWK) with tax funds on the basis of the budget approved by the Saxon State Parliament.
   The PIMC calculations have been carried out on a Bull Cluster at the Center for Information Services and High Performance Computing (ZIH) at Technische Universit\"at Dresden, and at the Norddeutscher Verbung f\"ur Hoch- und H\"ochstleistungsrechnen (HLRN) under grant \emph{shp00026}.

\section*{References}

\bibliographystyle{unsrt}
\bibliography{bibliography}

\end{document}